\documentclass[12pt]{article}

\newcommand{\be}{\begin{equation}}
\newcommand{\ee}{\end{equation}}
\newcommand{\bi}[1]{\vspace{-3mm} \bibitem{#1}}
\usepackage{epsfig,amsmath,amssymb,graphics,graphicx} 

\usepackage{graphics,graphicx}
\usepackage{amssymb,amsmath}
\usepackage{hyperref}
\usepackage{amsfonts}
\usepackage{epsfig}

\begin{document}
\begin{center}

{\it Mechanics of Materials. Vol.70. No.1. (2014) 106-114.}
\vskip 3mm

{\bf \large Lattice Model of Fractional Gradient and Integral Elasticity:} \\
\vskip 3mm
{\bf \large Long-Range Interaction of Gr\"unwald-Letnikov-Riesz Type} \\

\vskip 3mm
{\bf \large Vasily E. Tarasov} \\
\vskip 3mm

{\it Skobeltsyn Institute of Nuclear Physics,\\ 
Lomonosov Moscow State University, Moscow 119991, Russia} \\
{E-mail: tarasov@theory.sinp.msu.ru} \\

\begin{abstract}
Lattice model with long-range interaction of power-law type 
that is connected with difference of non-integer 
order is suggested.
The continuous limit maps the equations of motion of lattice particles
into continuum equations with fractional Gr\"unwald-Letnikov-Riesz derivatives.
The suggested continuum equations describe 
fractional generalizations of the gradient and integral elasticity.
The proposed type of long-range interaction allows us to have 
united approach to describe of lattice models for
the fractional gradient and fractional integral elasticity. 
Additional important advantages of this approach are the following: 
(1) It is possible to use this model of long-range interaction in numerical simulations since 
this type of interactions and the Gr\"unwald-Letnikov derivatives are 
defined by generalized finite difference;
(2) The suggested model of long-range interaction leads 
to an equation containing the sum of the Gr\"unwald-Letnikov derivatives, 
which is equal the Riesz's derivative. This fact allows us to get particular
analytical solutions of fractional elasticity equations.
\end{abstract}

\end{center}

\noindent
PACS: 45.10.Hj; 61.50.Ah; 62.20.Dc \\

\section{Introduction}

Discrete system of long-range interacting particles 
serve as a model for numerous applications in mechanics and physics 
\cite{Dyson1,Dyson3,CMP,NakTak2,NakTak4,S,CDR,LLI-2,LLI-3,LLI-4,LLI-6}. 
Long-range interactions are important type of interactions 
for complex media with non-local properties 
(see references in \cite{TarasovSpringer}). 

Using the fractional calculus \cite{SKM,KST}, 
we consider long-range interaction of a special type to describe 
a fractional generalization of elasticity theory.
We transform the set of equations of motion of coupled individual 
lattice particles into the equation of non-local continuum 
that contains spatial derivatives of non-integer orders.
It allows us to consider different lattice models 
for generalization of elasticity theory
by applying methods of fractional calculus.
The theory of fractional derivatives and integrals has wide applications 
\cite{CM,Hilfer,LA,KLM,Mainardi,TarasovRCD,Uchaikin3,IJMPB2013,TarasovSpringer}
and it is a powerful tool for the analysis of different nonlocal continuum models
with nonlocality of power-law type.

Non-local continuum mechanics has been treated with two different approaches:
the gradient models (weak non-locality) and 
the integral models (strong non-locality).
The correspondent constitutive relations have the form
\[
\sigma_{ij} = C_{ijkl} (\varepsilon_{kl} \pm l^2_s {\cal L} \, \varepsilon_{kl}) ,
\]
where $C_{ijkl}$ is the elasticity tensor, $\varepsilon_{ij}$  and $\sigma_{ij}$ are the strain and stress tensors, 
respectively. 
The additional parameter $l_s$ is an internal length scale
that can be defined by lattice constant.
The operator ${\cal L}$ is the integral or differential operator 
of integer order \cite{EE1972,AMPB,AS2002,ASS2002,AA2011}.
Recently, the problems of non-local continuum are described by means of fractional calculus.
Fractional models of integral non-local elasticity 
are considered in \cite{Laz,CCS-1,CCS-2,CCS-3,CPZ-1,CPZ-2,CPZ-3}.
In fractional elasticity models, the operator ${\cal L}$ 
is the integral or differential operator of non-integer order. 
To generalize continuum equations by using fractional calculus, 
we should represent these equation through 
the dimensionless coordinate variables.
Therefore the coordinates $x$, $y$, $z$, 
the vector ${\bf r}$, the value $r=|{\bf r}|$,
and the parameter $l^2_s$ are dimensionless in 
the suggested fractional models. 

In this paper we consider one-dimensional lattice model
with long-range interactions of of Gr\"unwald-Letnikov-Riesz type \cite{TarasovSpringer}. 
A feature of suggested long-range interactions 
is that the interactions terms have the form
of fractional differences of non-integer orders. 
The first advantage of this approach is based on the properties of 
the Gr\"unwald-Letnikov fractional derivative \cite{SKM,KST}.
The suggested type of long-range interaction of lattice particles
allows us to have united description of lattice models for
the fractional gradient and fractional integral elasticity that is
characterized by the non-locality of power-law type. 
The second important advantage of suggested approach is the ability to directly use 
the model of long-range interaction in numerical simulations since 
this type of interactions and the Gr\"unwald-Letnikov derivatives are 
defined by generalized difference of non-integer order. 
We assume that the suggested form of long-range interaction can be 
used for different scheme of simulations in fractional 
gradient and integral elasticity models. 

There are some problems with application of the Gr\"unwald-Letnikov derivatives, 
since there are currently very limited number of analytical solutions 
for differential equations with fractional Gr\"unwald-Letnikov derivatives 
in contrast to the equations with derivatives of Riemann-Liouville, Caputo and Riesz types \cite{KST}.
We note that the suggested model of long-range interaction leads 
to an equation containing the sum of the Gr\"unwald-Letnikov derivatives, 
which is equal the Riesz derivative.
Using this connection, we derive some particular analytical solutions \cite{KST} 
for fractional elasticity equations with the Riesz derivatives of non-integer orders.


\section{Gr\"unwald-Letnikov fractional derivatives and integrals}


\subsection*{Fractional differences of non-integer orders}

The Gr\"unwald-Letnikov derivatives have been 
introduced by Gr\"unwald in 1867 and independently by Letnikov in 1868. 
Definition of the Gr\"unwald-Letnikov fractional derivatives are based on 
a generalization of the usual differentiation of a function $f(x)$ 
of integer order $n$ of the forms
\[ D^n_x f(x) = \lim_{h \to  0} \frac{\Delta^n_h f(x)}{h^n} , \]
\[ D^n_x f(x) = \lim_{h \to  0} \frac{\nabla^n_h f(x)}{h^n} , \]
where $\Delta^n_h$ and $\nabla^n_h$ are forward and backward finite differences 
of order $n$ of a function $f(x)$ with a step $h$ and centered at the point $x$. 
The $n$th-order forward and backward differences are respectively given by
\be \label{dif-1}
\Delta^n_h f(x) = \sum_{k = 0}^{n} (-1)^k \binom{n}{k} f(x + (n -k) h),
\ee
\be \label{dif-2}
 \nabla^n_h f(x) = \sum_{k = 0}^{n} (-1)^k \binom{n}{k} f(x - k h) .
\ee
Note that
\be
\Delta^n_h f(x) = (-1)^n \nabla^n_{-h} f(x) .
\ee
The difference of a fractional order $\alpha>0$ is defined by 
the infinite series (see Section 20 in \cite{SKM})
\be \label{dif-3}
\nabla^{\alpha}_h f(x) = \sum_{k = 0}^{\infty} (-1)^k \binom{\alpha}{k} f(x - k h) ,
\ee
where the binomial coefficients are
\[ \binom{\alpha}{k} = 
\frac{\Gamma(\alpha+1)}{\Gamma(k+1) \Gamma (\alpha-k+1)} . \]
For  $h > 0$, the difference (\ref{dif-3}) 
is called left-sided fractional difference, 
and for $h < 0$ it is called a right-sided fractional difference.
We note that the series in (\ref{dif-3}) 
converges absolutely and uniformly 
for every bounded function $f(x)$ and $\alpha>0$. 

For the fractional difference, the semigroup property
\be \label{SGP-FD}
\nabla^{\alpha}_h  \nabla^{\beta}_h f(x) =
\nabla^{\alpha+\beta}_h f(x) ,\quad (\alpha>0 , \quad \beta>0) 
\ee
is valid for any bounded function $f(x)$ 
(see Property 2.29 in \cite{KST} page 121).

The Fourier transform of the fractional difference is given by
\[ {\cal F} \{\nabla^{\alpha}_h f(x) \}(k)= 
(1-\exp\{ikh\})^{\alpha}  {\cal F} \{f(x) \}(k)  \]
for any function $f(x) \in L_1(\mathbb{R})$ (see Property in \cite{KST} page 121).


\subsection*{Gr\"unwald-Letnikov fractional derivatives}

The definitions (\ref{dif-1}) and (\ref{dif-2}) are used 
to define the Gr\"unwald-Letnikov fractional derivatives 
by replacing $n \in \mathbb{N}$ in by $\alpha>0$. 
The value $h^n$ is replaced by $h^{\alpha}$, 
while the finite difference $\nabla^n_h$ is replaced by 
the difference $\nabla^{\alpha}_h$ of a fractional order $\alpha$. 

The left- and right-sided Gr\"unwald-Letnikov derivatives 
of order $\alpha>0$ are defined by
\be \label{GLD}
^{GL}D^{\alpha}_{x \pm} f(x) = 
\lim_{h \to  0+} \frac{\nabla^{\alpha}_{\pm h} f(x)}{|h|^{\alpha}} . 
\ee
Note that the Gr\"unwald-Letnikov derivatives for integer orders $\alpha=n \in \mathbb{N}$ is
\be
^{GL}D^{n}_{x \pm} f(x) = (-1)^n \, D^n_x .
\ee
We also note that these derivatives coincide with 
the Marchaud fractional derivatives of order $\alpha>0$
for $f(x)\in L_p(\mathbb{R})$, $1\leqslant p <\infty$ (see Theorem 20.4 in \cite{SKM}).
The properties of the Gr\"unwald-Letnikov fractional derivatives 
is described in Section 20 of the book \cite{SKM}.
Then (\ref{GLD}) can be represented by the Marchaud fractional derivatives 
\[ ^{GL}D^{\alpha}_{x \pm} f(x) =  
\frac{\alpha}{ \Gamma(1-\alpha)} 
\int^{\infty}_0  \frac{f(x)-f(x \mp z)}{z^{\alpha+1}} dz  \]
if $f(x) \in L_p(\mathbb{R})$, 
where $1<p<1/\alpha$ and $0<\alpha<1$.


\subsection*{Gr\"unwald-Letnikov fractional integral}

It is interesting that series (\ref{dif-3}) can be used for $\alpha<0$ 
(see Section 20 in \cite{SKM})
and equation (\ref{GLD}) defines the Gr\"unwald-Letnikov fractional integral if 
\be 
|f(x)|  < c (1+|x|)^{-\mu} , \quad \mu> |\alpha| . 
\ee
The existence of the Gr\"unwald-Letnikov fractional integral means that 
we have a united definition of fractional derivatives and integrals.
It allows us to have a united approach
to the fractional gradient and integral elasticity. 


\subsection*{Commutativity and associativity of Gr\"unwald-Letnikov derivatives.}

In \cite{ORT} has been showed that, for analytic functions,
the fractional derivatives of the Gr\"unwald-Letnikov type
have some nice and useful properties and a semi-group structure.
It is connected with the fact that differences of fractional order
satisfy the semigroup property (\ref{SGP-FD}).
Using the semi-group property 
\be \label{SGP}
^{GL}D^{\alpha}_{x \pm} \, ^{GL}D^{\beta}_{x \pm} = 
\, ^{GL}D^{\alpha+\beta}_{x \pm} ,
\quad (\alpha>0 , \quad \beta>0) .
\ee
Ortigueira, Rivero, Trujillo proved \cite{ORT} that the 
Gr\"unwald-Letnikov derivatives 
are commutative and associative operators.
These properties are very important to application 
in mechanics, but only a few types of 
fractional derivatives have them.
These properties allow us to represent 
a single fractional derivative, which appears 
in the equation of the lattice model, 
as a product of derivatives, 
and thus get the fractional gradient and integral elastic constitutive relations.


\subsection*{Gr\"unwald-Letnikov-Riesz fractional derivative}

We can define a fractional derivative of order $\alpha>0$ by
\be \label{GLR-der}
^{GLR}D^{\alpha}_{x} f(x) = \frac{1}{2\cos (\alpha \pi /2)}
\lim_{h \to  0+} 
\frac{ \nabla^{\alpha}_{h} f(x)+\nabla^{\alpha}_{-h} f(x)}{|h|^{\alpha}} . 
\ee
This derivative coincide (see Section 20.1 in \cite{SKM}) 
with the Riesz fractional derivative $ \, ^R{\bf D}^{\alpha}_x$ 
of order $\alpha>0$ if $\alpha \ne 1,2,3,...$, 
i.e. we have the relation
\be\label{GLR=R}
^{GLR}D^{\alpha}_{x} f(x) = \, ^R{\bf D}^{\alpha}_x f(x) ,  
\ee
where
\be \label{dif-GLR}
^R{\bf D}^{\alpha}_x f(x)
= -\frac{\alpha}{2 \Gamma(1-\alpha) \cos (\alpha \pi/2)}
\int^{\infty}_0 \frac{f(x+z)-2 f(x)+f(x-z)}{z^{\alpha+1}} dz  . 
\ee
Therefore the fractional derivative (\ref{GLR-der}) is called 
the Gr\"unwald-Letnikov-Riesz derivatives of order $\alpha>0$ ($\alpha \ne 1,2,3,...$) in \cite{SKM}.


\section{Lattice with linear nearest-neighbor interaction}

Let us consider
equations of motion for particles with the nearest-neighbor interaction of the form 
\be \label{2-CEM}
M \, \frac{\partial^2}{\partial t^2} u_n(t) = K \hat A^k_n u_k (t) + F_n (t) , \ee
where
\[ \hat A^m_n u_m (t) = \sum^{m=1}_{m=0} (-1)^{m+1} \, \Bigl( u_{n+m}(t)+u_{n-m}(t) \Bigr) . \]
For these equations we can give the well-known 
statement regarding the nearest-neighbor interaction 
(see for example \cite{Maslov,LLI-1,LLI-2}).

{\bf Proposition 1.} 
{\it In the continuous limit the lattice equations of motion (\ref{2-CEM})
maps into the continuum equation
\be \label{2-CME0}
\frac{\partial^2 u(x,t)}{\partial t^2} =
C^2_e \, \frac{\partial^2 u(x,t)}{\partial x^2} + \frac{1}{\rho} f(x,t) ,
\ee
with the mass density $\rho= M/ A h$, 
the Young's modulus $E=K h/A$, the force density 
$f(x,t)=F(x,t)/A h$, 
the cross-section area of the medium $A$ and 
the inter-particle distance $h$, where
\be \label{C2e}
C^2_e =\frac{E}{\rho}=\frac{K \, h^2}{M} 
\ee
is a finite parameter.}

{\bf Proof}.
To derive the equation for the field $\hat u(k,t)$, we
multiply equation (\ref{2-CEM}) by $\exp(-ikn \Delta x)$, where $\Delta x = h$
and summing over $n$ from $-\infty$ to $+\infty$. Then
\be \label{2-DD1}
M \, \sum^{+\infty}_{n=-\infty} e^{-ikn \Delta x} 
\frac{\partial^2 u_n}{\partial t^2} =
K \, \sum^{+\infty}_{n=-\infty} \
e^{-ikn \Delta x}  [u_{n+1}-2u_n+u_{n-1}] +
\sum^{+\infty}_{n=-\infty} e^{-ikn\Delta x} F(n,t) .
\ee
The first term on the right-hand side of (\ref{2-DD1}) is
\[
\sum^{+\infty}_{n=-\infty} \
e^{-ikn \Delta x}  \, \Bigl( u_{n+1}-2u_n+u_{n-1} \Bigr) 
= \]
\[ =\sum^{+\infty}_{n=-\infty} \
e^{-ikn \Delta x}  u_{n+1} -
2 \sum^{+\infty}_{n=-\infty} \
e^{-ikn \Delta x}  u_n +
\sum^{+\infty}_{n=-\infty} \
e^{-ikn \Delta x}  u_{n-1}= \]
\[ =e^{ik\Delta x} 
\sum^{+\infty}_{m=-\infty} \
e^{-ik m \Delta x}  u_{m} -
2 \sum^{+\infty}_{n=-\infty} \
e^{-ikn \Delta x}  u_n +
e^{-ik \Delta x} 
\sum^{+\infty}_{s=-\infty} \
e^{-ik s \Delta x}  u_{s} =
\]
\[ = e^{ik\Delta x} \hat{u}(k,t)- 2 \hat{u}(k,t) +
e^{-ik \Delta x} \hat{u}(k,t)= \]
\[ =[e^{ik\Delta x} +e^{-ik \Delta x}-2] \hat{u}(k,t)= 
2\Bigl( \cos \left( k \Delta x \right)-1 \Bigr) \hat{u}(k,t) =
-4 \sin^2 \left( \frac{k \Delta x}{2} \right) \hat{u}(k,t) . \]
Here we use the definition of $\hat{u}(k,t)$
on $[-k_0 /2, k_0 /2]$ by the equation 
\be \label{ukt}
\hat{u}(k,t) = \sum_{n=-\infty}^{+\infty} \; u_n(t) \; e^{-i k x_n} =
{\cal F}_{\Delta} \{u_n(t)\} ,
\ee
and $x_n = n \, h$, where $h = 2\pi/k_0$ is distance between equilibrium 
positions of the lattice particles.

As a result, we have
\be \label{2-simple}
M \, \frac{\partial^2 \hat{u}(k,t)}{\partial t^2} = 
K  \; \hat{J}_{\alpha}(k \Delta x) \; \hat{u}(k,t) +
\mathcal{F}_{\Delta} \{ F_n (t) \} ,
\ee
where
\be \label{2-J2}
\hat{J}_{\alpha} (k \Delta x)= -4 \sin^2 \left( \frac{k \Delta x}{2} \right) .
\ee
For $h=\Delta x \to 0$, the asymptotic behavior of the sine is
$\sin(k \Delta x / 2 ) \sim k \Delta x/2$. 
Then (\ref{2-J2}) can be represented by
\[ \hat{J}_{\alpha} (k \Delta x) \approx  -  \left( k \Delta x \right)^2 . \]
Using the finite parameter $C^2_e=K \, h^2/M$, 
after division by the cross-section area of the medium $A$ and the inter-particle distance $h$,
the transition to the limit $h=\Delta x \to 0$ in equation (\ref{2-simple}) gives 
\be \label{2-DD2}
\frac{\partial^2  \tilde u(k,t)}{\partial t^2}=
- C^2_e \, k^2 \tilde u(k,t) + \frac{1}{\rho} {\cal F} \{f(x,t)\} ,
\ee
where we use $0<|C^2_e|<\infty$.
The inverse Fourier transform ${\cal F}^{-1}$ of (\ref{2-DD2}) has the form
\[ \frac{\partial^2 {\cal F}^{-1}\{ \tilde u(k,t)\} }{\partial t^2}=
- C^2_e \, {\cal F}^{-1} \{k^2 \tilde u(k,t)\} + \frac{1}{\rho} \, f(x,t) . \]
Then we can use the connection between second derivative and 
its Fourier transform in the form 
$k^2 \longleftrightarrow - \partial^2/\partial x^2$. 
As a result, we obtain the continuum equation (\ref{2-CME0}). 
$\ \ \ \Box$ \\

As a result, we prove that equations (\ref{2-CEM}) give the continuum equation 
with derivatives of second order only.


\section{Lattice with Gr\"unwald-Letnikov-Riesz long-range interaction}

In this section we describe the type of long-range interaction 
that is suggested in \cite{TarasovSpringer} (see Section 8.19). 
Let us consider a lattice system of interacting particles,
whose displacements from the equilibrium are $u_n(t)$, 
where $n \in \mathbb{Z}$.  
We assume that the system is described by the equations of motion
\be \label{Eq-GL1}
M \, \frac{\partial^2}{\partial t^2} u_n(t) + 
g \, \hat B^m_n (\alpha) u_m(t) - F_n(t) =0 , \ee
where $g$ is the coupling constant 
for long-range interaction that have the form
\be \label{LuGamma}
\hat B^m_n (\alpha) u_m(t)= \sum^{+\infty}_{m=0} b_{\alpha}(m) \, \Bigl( u_{n+m}(t)+u_{n-m}(t) \Bigr) ,
\ee
and the function $b_{\alpha}(m)$ is
\be \label{JnmGamma} 
b_{\alpha}(m)= \frac{(-1)^m}{\Gamma(m+1) 
\, \Gamma(\alpha -m+1)} .
\ee
This type of long-range interaction has been called 
the Gr\"unwald-Letnikov-Riesz interaction 
(see Section 8.19 in \cite{TarasovSpringer}). 
Let us give the main statement regarding this interaction. 

{\bf Proposition 2.}
{\it In the limit $h\to  0$ lattice equations (\ref{Eq-GL1}),
(\ref{LuGamma}) with (\ref{JnmGamma})
give the continuum equation
\be \label{Eq-GL2}
\frac{\partial^2 u(x,t)}{\partial t^2} + C(\alpha) \, 
^{GLR}D^{\alpha}_{x}  \, u(x,t)  - \frac{1}{\rho} f(x,t) =0 , \quad 
(\alpha \in \mathbb{R}, \alpha \ne \pm 1, \pm 3, \pm 5, ...)  , 
\ee
where
\be \label{C-alpha}
C(\alpha) =  \frac{ 2 \cos(\alpha \, \pi/2) \, 
g \, h^{\alpha}}{\Gamma(\alpha+1) \, M}
\ee
is a finite parameter, and
\be
^{GLR}D^{\alpha}_{x} = \frac{1}{2 \, \cos (\alpha \pi/ 2)} \, 
\Bigl(\, ^{GL}D^{\alpha}_{x+} + \, ^{GL}D^{\alpha}_{x-} \Bigr) 
\ee
is the Gr\"unwald-Letnikov-Riesz fractional derivative of order $\alpha$,
and $u(x,t)$ is a smooth function such that $u(nh,t)=u_n(t)$.}


{\bf Proof.}
We define smooth functions $u(x,t)$ and $F(x,t)$ such that
\[ u(nh,t)=u_n(t) , \quad F(nh,t)=F_n(t)  . \]
Then equations (\ref{Eq-GL1}), (\ref{LuGamma}) with (\ref{JnmGamma}) 
can be represented as
\be \label{Eq-GL4}
\frac{\partial^2 u(x,t)}{\partial t^2}+ \frac{g \, h^{\alpha}}{M}
\sum^{+\infty}_{m=0} b_{\alpha}(m) \, \frac{1}{h^{\alpha}} \, 
\Bigl( u(x+mh,t)+u(x-mh,t)\Bigr) - \frac{1}{M} F(x,t) =0. \ee
After division by the cross-section area of the medium $A$ 
and the inter-particle distance $h$ it is found that
\be \label{Eq-GL4b}
\frac{\partial^2 u(x,t)}{\partial t^2}+
\frac{C(\alpha)}{2 \, \cos(\alpha \pi /2)} \, 
\sum^{+\infty}_{m=0} 
\frac{(-1)^m \, \Gamma(\alpha+1)}{\Gamma(\alpha -m+1)\, \Gamma(m+1) } \, \frac{u(x+mh,t)+u(x-mh,t)}{h^{\alpha}} - 
\frac{1}{\rho} f(x,t) =0 \ee
with the mass density $\rho= M/ A h$, 
the force density $f(x,t)=F(x,t)/A h$, and 
$C(\alpha)$ is defined by (\ref{C-alpha}).
Using the definitions of the left-sided and 
right-sided fractional differences and 
the limit $h \to 0+$, we obtain 
\be \label{Eq-GL5}
\frac{\partial^2 u(x,t)}{\partial t^2} + \frac{C(\alpha)}{2 \, \cos(\alpha \pi /2)}  \, 
\lim_{h \to  0}  \, 
\frac{ \nabla^{\alpha}_{h} u(x,t)+\nabla^{\alpha}_{-h} u(x,t)}{|h|^{\alpha}} - \frac{1}{\rho} f(x,t) =0  . 
\ee
Using the Gr\"unwald-Letnikov-Riesz derivative (\ref{GLR-der}), 
equation (\ref{Eq-GL5}) can be rewritten in the form (\ref{Eq-GL2}).
$\ \ \ \Box$ \\


\section{Fractional gradient and integral elasticity of Gr\"unwald-Letnikov-Riesz type}

Let us consider a system of interacting particles,
whose displacements from the equilibrium are $u_n(t)$, where $n \in \mathbb{Z}$.  
We assume that the system is described by the equations of motion
\be \label{FGIE-1}
M \, \frac{\partial^2 u_n(t)}{\partial t^2}  -
K \, \hat A^m_n u_m(t)  + g \hat B^m_n(\alpha) u_m(t) - F_n(t) = 0 . \ee
In the limit $h \to 0$ equations (\ref{FGIE-1}) gives the continuous medium equation
\be \label{FGIE-2}
\frac{\partial^2 u(x,t)}{\partial t^2} -
C^2_e \, \frac{\partial^2 u(x,t)}{\partial x^2} +  C(\alpha)  \, 
^{GLR}D^{\alpha}_{x} \, u(x,t)  - \frac{1}{\rho} f(x,t) =0 , 
\ee 
where $C^2_e$ is defined by (\ref{C2e}), 
$C(\alpha)$ is defined by (\ref{C-alpha}), and
\[ \alpha \in \mathbb{R}, \quad \alpha \ne \pm 1, \pm 3, \pm 5, ...  \ .  \]


Let us consider equation (\ref{FGIE-2}) for two cases:
$\alpha >2$ and $0<\alpha <2$.

For $\alpha>2$, we can use
$\, ^{GL}D^{1}_{x \pm} = \pm D^1_x$, 
and the semi-group property (\ref{SGP}), to represent 
the Gr\"unwald-Letnikov derivatives in the form
\be \label{DDD}
^{GL}D^{\alpha}_{x \pm} = 
D^1_x \ ^{GL}D^{\alpha-2}_{x \pm} \, D^1_x  \quad (\alpha >2).
\ee
Therefore, we have
\be
^{GL}D^{\alpha}_{x+} + \, ^{GL}D^{\alpha}_{x-}  =
D^1_x \Bigl(\, ^{GL}D^{\alpha-2}_{x+} + \, ^{GL}D^{\alpha-2}_{x-} \Bigr) \, D^1_x .
\ee
As a result, we can use
\be \label{DDD-2}
^{GLR}D^{\alpha}_{x} = D^1_x \ ^{GLR}D^{\alpha-2}_{x} \, D^1_x .
\ee
in order to rewrite equation (\ref{FGIE-2}) in the form
\be \label{Eq-GL2b}
\frac{\partial^2 u(x,t)}{\partial t^2} - C^2_e \, D^2_x u(x,t) 
+ C(\alpha) \, D^1_x \, ^{GLR}D^{\alpha-2}_{x} \, D^1_x u(x,t)
- \frac{1}{\rho} f(x,t) =0 . 
\ee
The correspondent constitutive relation 
for the continuum equation (\ref{Eq-GL2b})
can be derived by the momentum balance equation
\be \label{MBE}
\rho \,  \frac{\partial^2 u(x,t)}{\partial t^2} = D^1_x \sigma (x) + f(x,t) ,
\ee
and the strain-displacement relation for small deformations 
\be \label{1.2}
\varepsilon (x)  = D^1_x u(x) .
\ee
The commutative and associative properties of fractional Gr\"unwald-Letnikov derivative
and relation (\ref{DDD}) allow us to represent a single fractional derivative in equation (\ref{Eq-GL2}), 
as a product of derivatives, 
and thus get the fractional gradient constitutive relations.

Using (\ref{MBE}) and (\ref{1.2}), equation (\ref{Eq-GL2b}) gives
the fractional constitutive relation in the following form
\be \label{2.2a} 
\sigma(x,t) = E \, \left( \varepsilon (x,t) \mp
l^2_s (\alpha) \ ^R{\bf D}^{\alpha-2}_{x} \varepsilon (x,t) \right)
\quad (\alpha >2),
\ee
where $E$ is the Young's modulus,
$ ^R{\bf D}^{\alpha-2}_x$ is the Riesz fractional derivative of order $\alpha -2$, which 
is equivalent by (\ref{GLR=R}) to the 
Gr\"unwald-Letnikov-Riesz fractional derivative 
$\, ^{GLR}D^{\alpha-2}_{x}$. 
As a result, we can state that equation (\ref{FGIE-2}) 
with $\alpha >2$ describes a fractional gradient elasticity.
In equation (\ref{2.2a}), we use
\be \label{lalpha}
l^2_s(\alpha) = \frac{|C(\alpha)| \, \rho}{E} = \frac{2 \, |g| \, h^{\alpha-2}}{K} \frac{|\cos(\alpha \pi /2)|}{\Gamma(\alpha+1)}
\ee
is the scale parameter of fractional elasticity. 
Sign in front of this scale parameter $l^2_s(\alpha)$ in equations (\ref{2.2a}) 
is determined by the sign of the expression $g \, \cos (\alpha \pi /2)$,
i.e. the sign of the coupling constant $g$ of lattice vibrations and the value of the
order $\alpha$ of long-rang interactions.
If $g \, \cos (\alpha \pi /2) >0$, then we get the minus 
in front of $l^2_s(\alpha)$ in (\ref{lalpha}).
For the case $\alpha=4$ and $g>0$, we derive 
the constitutive relations for the gradient elasticity model 
with the minus in front of $l^2_s$. 
We get the equations for phenomenological gradient elasticity model from the lattice model equations. 
Note that normally it is considered a phenomenological model 
does not have a corresponding microscopic model \cite{AA2011}.


For the case $0<\alpha<2$, we cannot use 
the properties (\ref{DDD}) and (\ref{DDD-2}) 
because semi-group relation (\ref{SGP}) holds for positive orders.
For $0<\alpha<2$, we can use the relation (\ref{GLR=R})
and the Riesz's analytic continuation 
(see equation (17) in \cite{Riesz})
of fractional integrals of order $\alpha$ 
to negative values of orders $\alpha > -p$ in the form
\be
^R{\bf I}^{\alpha} =
(-1)^p \, ^R{\bf I}^{\alpha + 2p} \, \Delta^p ,
\ee
where $p \in \mathbb{N}$ and $\, ^R{\bf I}^{\alpha + 2p}$ 
is the Riesz fractional integral of order $\alpha + 2p$. 
Using equations (17), (21) and (22) from 
the Riesz's review paper \cite{Riesz}, we get
\be
^R{\bf D}^{\alpha}_{x} = - \, ^R{\bf I}^{2-\alpha} \, \Delta =
\Delta \, ^R{\bf I}^{4-\alpha} \, \Delta .
\ee
For one-dimensional case, we have 
$\Delta =(D^1_x)^2$, and we can use the relation
\be \label{Int-J}
\ ^R{\bf D}^{\alpha}_{x} = \Delta \, ^R{\bf I}^{4-\alpha} \, \Delta = D^1_x \, {\bf J}^{2-\alpha}_x \, D^1_x , 
\ee
where ${\bf J}^{2-\alpha}_x$ is the fractional integral 
operator of order $(2-\alpha)$ that is defined by  
\be \label{J-2-a}
{\bf J}^{2-\alpha}_x = D^1_x \, ^R{\bf I}^{4-\alpha}_x \, D^1_x 
\ee

As a result, equation (\ref{FGIE-2}) with $\alpha <2$ describes 
a fractional integral elasticity.
Using (\ref{MBE}), (\ref{1.2}) and (\ref{Int-J}), equation (\ref{FGIE-2}) gives
the fractional constitutive relation in the following form
\be \label{2.2b} 
\sigma(x,t) = E \, \left( \varepsilon (x,t) \mp 
 l^2_s (\alpha) \ ^R{\bf J}^{2-\alpha}_{x} \varepsilon (x,t) \right) 
\quad (0< \alpha <2),
\ee
where ${\bf J}^{2- \alpha}_x $ is the fractional integral (\ref{J-2-a}) of order $0<2-\alpha<2$, and $l^2_s (\alpha)$ is defined by (\ref{lalpha}). Sign in front of this scale parameter $l^2_s(\alpha)$ in equations (\ref{2.2b}) 
is also determined by the sign of the expression $g \, \cos (\alpha \pi /2)$.


Let us give a remark about the scale parameter $l_s(\alpha)$.
Equation (\ref{lalpha}) can
lead to incorrect conclusion about the behavior 
of the parameter $l^2_s(\alpha)$ for $h \to 0$ in the case $0<\alpha<2$. 
Using (\ref{C2e}), the dimensionless parameter (\ref{lalpha}) 
can be written as
\be \label{lalpha2}
l^2_s(\alpha) =  \frac{2\, |g| \, h^{\alpha}}{C^2_e \, M} 
\frac{|\cos(\alpha \pi /2)|}{\Gamma(\alpha+1)} .
\ee
Because the value of $C^2_e$ if finite, then
behavior of the parameter $l^2_s(\alpha)$ for
$h \to 0$ has the identical type for $\alpha>2$ and $0<\alpha<2$, 
such that  $l^2_s(\alpha)$ is proportion to $h^{\alpha}$.


As a result, equations (\ref{2.2a}) and (\ref{2.2b}) 
describe the fractional gradient and fractional 
integral elasticity of non-local continuum.
If $0<\alpha <2$, we have a fractional integral elasticity, and
if $\alpha>2$, then equation describes 
fractional gradient elasticity.
It can call the fractional elasticity 
of Gr\"unwald-Letnikov-Riesz type.


\section{Solutions of fractional elasticity equations}

Let us consider more general lattice system of interacting particles
that is described by the equations of motion
\be \label{FGIE-1c}
M \, \frac{\partial^2 u_n(t)}{\partial t^2} + 
K \, \hat A^m_n u_m(t)  + 
\sum^N_{k=1} g_k \, \hat B^m_n(\alpha_k) u_m(t) = F_n(t) .  \ee
The correspondent continuum equation for the fractional elasticity 
of the Gr\"unwald-Letnikov-Riesz type has the form
\be \label{FGE-0}
\frac{\partial^2 u(x,t)}{\partial t^2} - C^2_e \, D^2_x u(x,t) +
\sum^N_{k=1} C(\alpha_k) \, ^R{\bf D}^{\alpha_k}_x u(x) = \frac{1}{\rho} f(x) .
\ee
Here to get analytical solution of the fractional equations of nonlocal elasticity,
we use the equivalence (see Section 20.1 in \cite{SKM}) 
of the Gr\"unwald-Letnikov-Riesz derivative $\, ^{GLR}D^{\alpha}_{x}$
and the Riesz derivative $ ^R{\bf D}^{\alpha}_x$ in the form
\be 
^{GLR}D^{\alpha}_{x} = \frac{1}{2\cos (\alpha \pi /2)} 
\Bigl(\, ^{GL}D^{\alpha}_{x+} + \, ^{GL}D^{\alpha}_{x-} \Bigr) =\, ^R{\bf D}^{\alpha}_x . 
\ee
Equation (\ref{FGE-0}) can be considered as a fractional generalization 
of the higher-order strain-gradient models \cite{AS2002} and 
the higher-order integral elasticity models.
In the static case ($D^2_t u(x,t)=0$) 
the fractional  elasticity equations (\ref{FGE-0}) is
\be \label{FGE}
\sum^N_{k=1} C(\alpha_k) \, ^R{\bf D}^{\alpha_k}_x u(x) - C^2_e \, D^2_x u(x) = \frac{1}{\rho} f(x) ,
\ee
where $x\in \mathbb{R}$, $N \in \mathbb{N}$ and $m \ne 1$, $C(\alpha_k) \in \mathbb{R}$, 
$\alpha_N> .... \alpha_1>0$

Equations (\ref{FGE-0}) and (\ref{FGE}) involve the one-dimensional Riesz fractional derivatives given by
\be
^R{\bf D}^{\alpha_k}_x u(x)= \frac{1}{d_1(l,\alpha_k)} 
\int^{\infty}_{-\infty} \frac{(\Delta^m_z u)(z)}{|z|^{\alpha_k+1}dz } ,
\ee 
where $\alpha_k < l$, $k=1,2,...,m$, 
and $(\Delta^m_z f)(z)$ is a finite difference of
order $m$ of a function $f(x)$ with a vector step $z \in \mathbb{R}$
and centered at the point $x \in \mathbb{R}$:
\[ (\Delta^m_z f)(z) =\sum^m_{k=0} (-1)^k \frac{m!}{k!(m-k)!} \, f(x-kz) . \]
The constant $d_n(m,\alpha)$ is defined by
\[ d_1(m,\alpha)=\frac{\pi^{3/2} A_m(\alpha)}{2^{\alpha} 
\Gamma(1+\alpha/2) \Gamma(1/2+\alpha/2) \sin (\pi \alpha/2)} , \]
where
\[ A_m(\alpha)=\sum^m_{j=0} (-1)^{j-1} \frac{m!}{j!(m-j)!} \, j^{\alpha} . \]
Note that the hypersingular integral ${\bf D}^{\alpha}_x f(x)$ does not 
depend on the choice of $m>\alpha$.

Equations (\ref{FGE}) are solvable, and it particular solutions are given 
(see Theorem 5.25 in \cite{KST}) by the formula
\be
u(x) = \frac{1}{\rho} \int^{+\infty}_{-\infty} G_{\alpha} (x-z) \, f(z) \, dz ,
\ee
where
\be
G_{\alpha} (x-z) =\frac{1}{\pi} \int^{\infty}_0 
\left( \sum^m_{k=1} C(\alpha_k) \lambda^{\alpha_k} + C^2_e \lambda^2 \right)^{-1} 
\cos(\lambda |x|) \, d \lambda .
\ee

Let us consider one-dimensional W. Thomson (1848) problem \cite{LL}.
We determine the deformation of an infinite elastic continuum,
when a force is applied to a small region in it.
We consider one-dimensional elastic media with power-law nonlocality that is described 
by the equation
\be \label{FGEe}
- C^2_e \, D^{2}_x u(x) + C(\alpha) \, ^R{\bf D}^{\alpha}_x u(x)  = \frac{1}{\rho} f(x) .
\ee
Note that equation (\ref{FGEe}) coincides with equation 
(\ref{FGIE-2}) for static case.
If we consider the deformation at positions $x$, which are larger compare with the size of the region,
where the force is applied, we can suppose that the force is applied at a point.
In this case, we have
\be \label{deltaf}
f(x) =f_0 \delta(x) . 
\ee
Then the deformation, which is a particular solution of equation (\ref{FGEe}), will be
described by the equation
\be \label{SOL-1}
u(x) = \frac{f_0}{\pi \rho } \int^{\infty}_0 
\frac{\cos(\lambda |x|)}{ C^2_e \, \lambda^2 +  C(\alpha) \lambda^{\alpha}} \, d\lambda .
\ee
If $0<\alpha <2$, the solution (\ref{SOL-1}) corresponds 
to the fractional integral elasticity, and
if $\alpha>2$, then the solution (\ref{SOL-1}) corresponds 
to the fractional gradient elasticity.


We can consider more general model of lattice with long-range interaction in $\mathbb{R}^3$, 
where all particles are displaced from its equilibrium in one direction, 
and the displacement of particles is described by a scalar field $u(x)$, where 
$x=|{\bf r}|$ and ${\bf r} \in \mathbb{R}^3$.
The correspondent continuum equation of the fractional elasticity model is
\be \label{FPDE-4-2b}
-\Delta u(x) + \frac{C(\alpha)}{C^2_e} ((-\Delta)^{\alpha/2} u) (x) = \frac{1}{C^2_e \, \rho} f(x) ,
\ee
where $(-\Delta)^{\alpha/2}$ is the fractional Laplacian \cite{KST}. 
The displacement vector $u(x)$ of the point force (\ref{deltaf}) has the following form
\be \label{Pot-2-2b}
u(x) = \frac{1}{4 \pi C^2_e \rho} \frac{f_0}{x} \, \cdot \, C_{2, \alpha} (x) ,
\ee
where $x=|{\bf r}|$, and
\be
C_{2, \alpha} (x) = \frac{2}{\pi} \int^{\infty}_0 \frac{ \lambda \, \sin (\lambda x)}{ 
\lambda^2+ (C(\alpha)/ C^2_e) \lambda^{\alpha}  } \, d \lambda.
\ee


We note that the asymptotic behavior $x=|{\bf r}| \to 0$ of the scalar field $u(x)$ 
does not depend on the parameter $\alpha$. 
Using (see equation (1) of Section 2.3 in the book \cite{BE}), 
we obtain the asymptotic behavior ($x \to \infty$) for $C_{2, \alpha}(x)$ with $\alpha<2$ in the form
\be
C_{2,\alpha} (x) = \frac{2}{\pi}
\int^{\infty}_0 \frac{\lambda \sin (\lambda x)}{\lambda^2+ (C(\alpha)/ C^2_e) \lambda^{\alpha}}  \, d \lambda \approx
A_0(\alpha) \frac{1}{x^{2-\alpha}} + \sum^{\infty}_{k=1} A_k(\alpha) \frac{1}{x^{(2-\alpha)(k+1)}} ,
\ee
where 
\be
A_0(\alpha)= \frac{2 C^2_e}{\pi C(\alpha)} \, \Gamma(2-\alpha) \, \sin \left( \frac{\pi}{2}\alpha \right) ,
\ee
\be
A_k (\alpha) = -\frac{2}{\pi} \left(\frac{C^2_e}{C(\alpha)}\right)^{k+1} 
\int^{\infty}_0  z^{(2-\alpha)(k+1)-1} \, \sin(z) \, dz .
\ee

As a result, we have in the framework the fractional elasticity, 
the displacement field of the point force in the 
infinite media with non-locality of power-law type is given by
\be
u(x) \ \approx \ \frac{A_0(\alpha)}{4 \pi C^2_e \rho} \,
\cdot \, \frac{f_0}{x^{3-\alpha}} \quad (0< \alpha<2)
\ee
for the long distance $x \gg 1$. 

In Figures 1-3 we present some plots of the factor $C_{2,0}(x) =\exp (-x)$ and 
factors $C_{2,\alpha}(x)$ with $C(\alpha)/ C^2_e=1$ 
for different orders of $0 < \alpha < 2$, i.e. for the fractional integral elasticity. 
The values of the factors $C_{2,\alpha}(x)$ and $C_{2,0}(x)$ 
are plotted along the Y-axis, and the values of 
the position $x=|{\bf r}|$ are plotted along the X-axis.


For the case $\alpha >2$, i.e. for the fractional gradient elasticity, 
the asymptotic behavior $x=|{\bf r}| \to \infty$ of $u(x)$ 
does not depend on the parameter $\alpha$.
The asymptotic behavior of the displacement field $u (|{\bf r}|)$ for $|{\bf r}| \to 0$ is given by
\be \label{Cab-2}
u (x) \ \approx \ 
\frac{f_0  \, \Gamma((3-\alpha)/2)}{2^{\alpha} \, \pi^2 \sqrt{\pi} \, \rho \, C(\alpha) \, \Gamma(\alpha/2)} \,
\cdot \,  x^{\alpha-1} , \quad (2<\alpha<3),
\ee
\be \label{Cab-3}
u (x) \ \approx \ 
\frac{f_0}{2 \pi \, \alpha \, \rho \, (C^2_e)^{1-3/\alpha} \, C^{3/\alpha}(\alpha) \, \sin (3 \pi / \alpha)}
, \quad (\alpha>3) .
\ee
Note that the function $C_{2,\alpha}$ for the fractional gradient case ($\alpha>2$)
has a maximum. 
If $\alpha =4$ then we have the well-known case of 
gradient elasticity \cite{AA2011}. 

In Figures 4-6 we present some plots of the factor $C_{2,0}(x) =\exp (-x)$ and 
factors $C_{2,\alpha}(x)$ with $C(\alpha)/ C^2_e=1$ 
for different orders of $2 < \alpha < 6$, i.e. for the fractional gradient elasticity. 
The values of the factors $C_{2,\alpha}(x)$ and $C_{2,0}(x)$ 
are plotted along the Y-axis, and the values of 
the position $x=|{\bf r}|$ are plotted along the X-axis.



\begin{figure}[H]
\begin{minipage}[h]{0.47\linewidth}
\resizebox{11cm}{!}{\includegraphics[angle=-90]{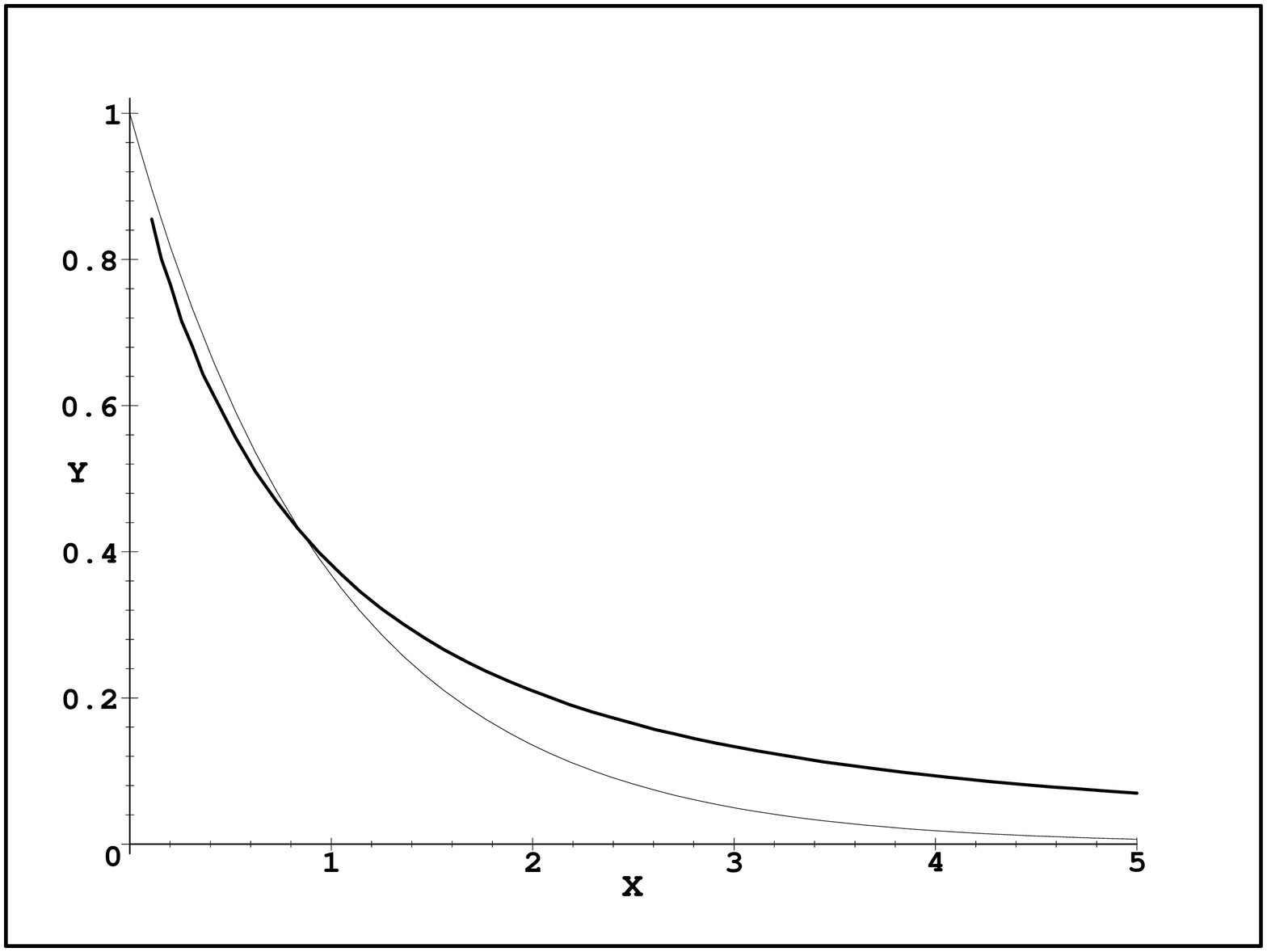}} 
\end{minipage}
\caption{Plot of the function $y=C_{2,\alpha}(x)$ with  
$\alpha=0.7$ is drawn with black color and 
the factor $y=C_{2,0}(x) =\exp (-x)$ is drawn by gray color,
where $x=|{\bf r}|$ and $C(\alpha)/ C^2_e=1$.}
\label{Plot1}
\end{figure}

\begin{figure}[H]
\begin{minipage}[h]{0.47\linewidth}
\resizebox{11cm}{!}{\includegraphics[angle=-90]{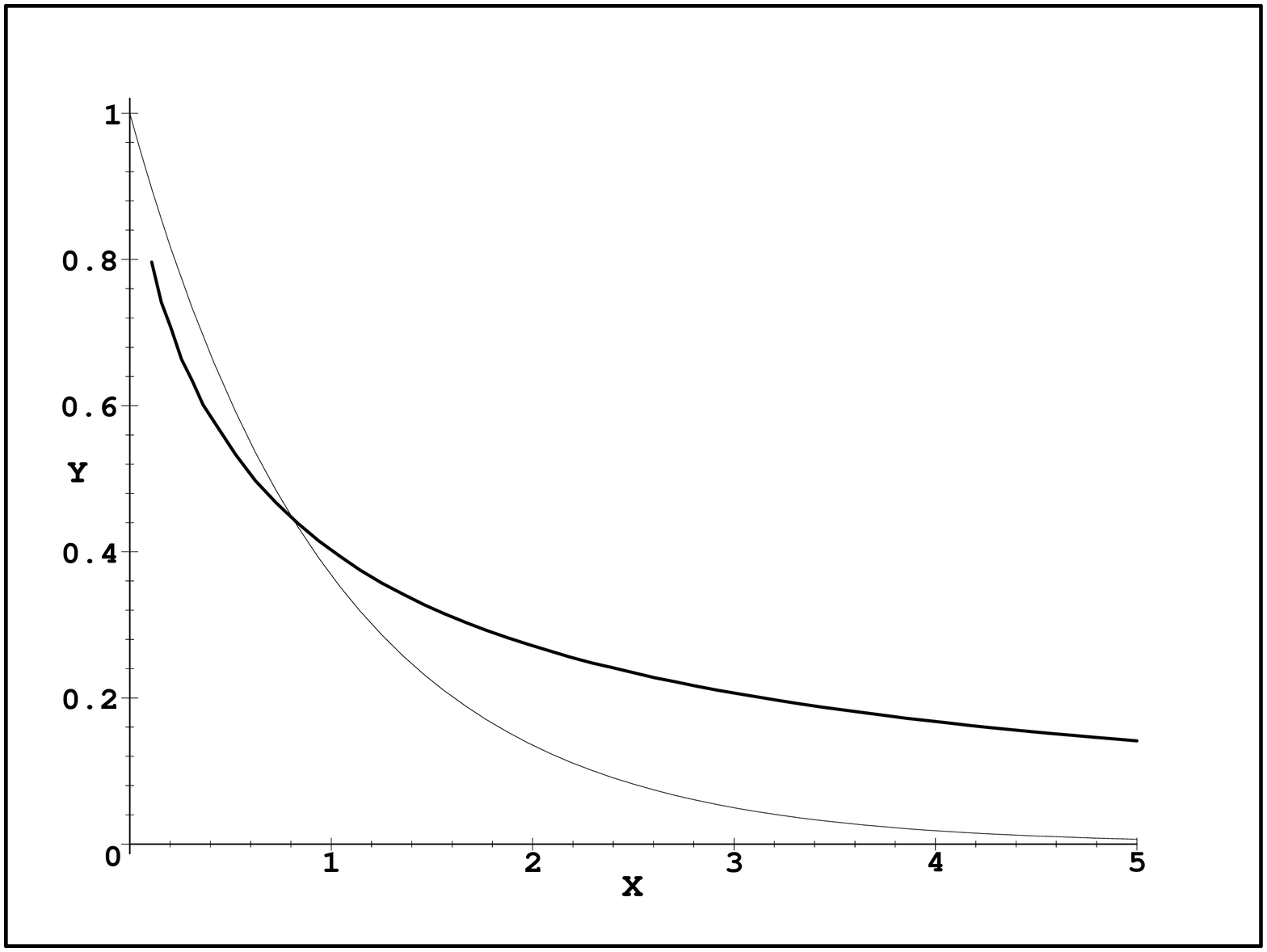}} 
\end{minipage}
\caption{Plot of the function $y=C_{2,\alpha}(x)$ with  
$\alpha=1.1$ is drawn with black color and 
the factor $y=C_{2,0}(x) =\exp (-x)$ is drawn by gray color,
where $x=|{\bf r}|$ and $C(\alpha)/ C^2_e=1$.}
\label{Plot2}
\end{figure}

\begin{figure}[H]
\begin{minipage}[h]{0.47\linewidth}
\resizebox{11cm}{!}{\includegraphics[angle=-90]{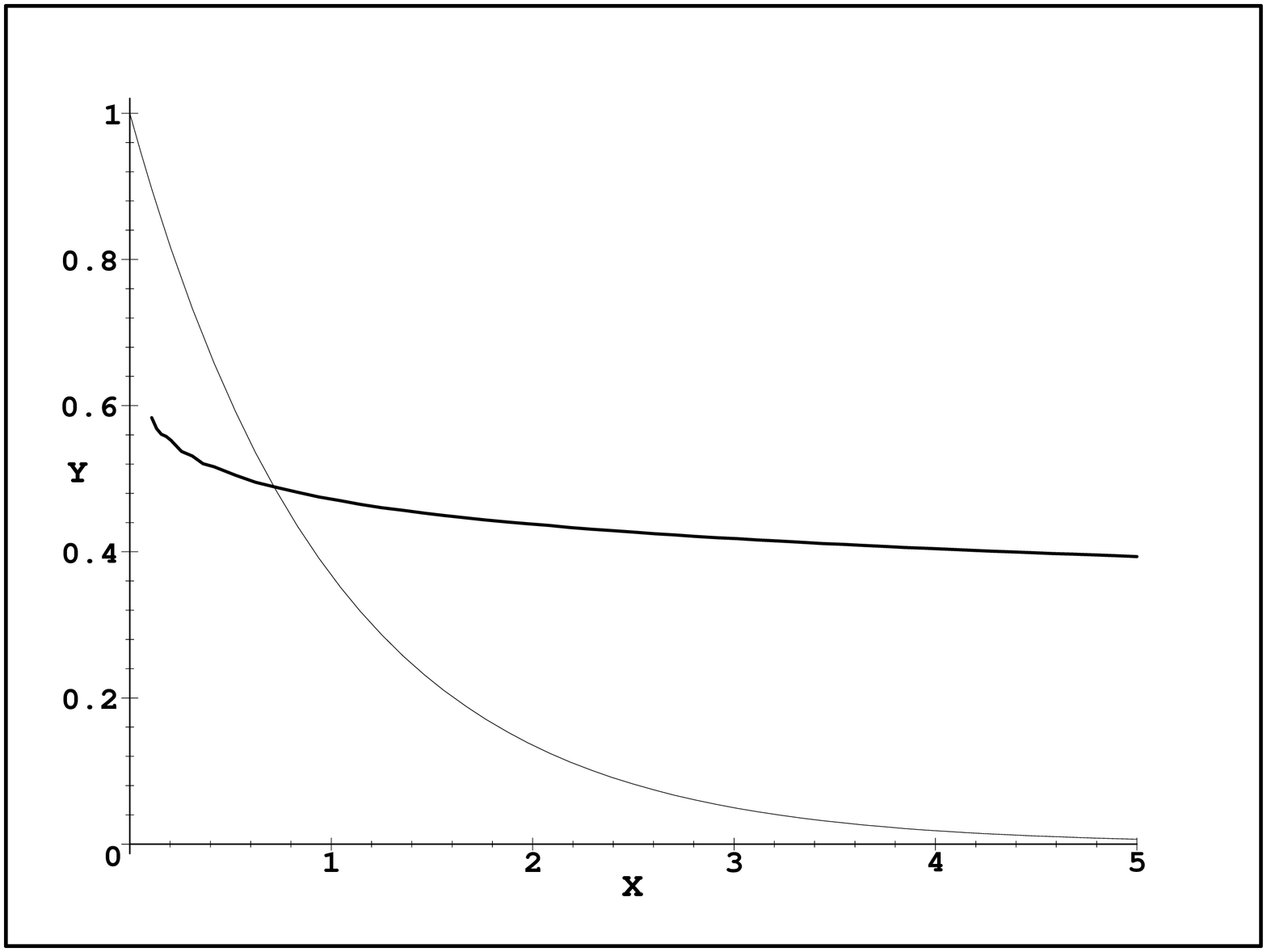}} 
\end{minipage}
\caption{Plot of the function $y=C_{2,\alpha}(x)$ with  
$\alpha=1.8$ is drawn with black color and 
the factor $y=C_{2,0}(x) =\exp (-x)$ is drawn by gray color,
where $x=|{\bf r}|$ and $C(\alpha)/ C^2_e=1$.}
\label{Plot3}
\end{figure}



\begin{figure}[H]
\begin{minipage}[h]{0.47\linewidth}
\resizebox{11cm}{!}{\includegraphics[angle=-90]{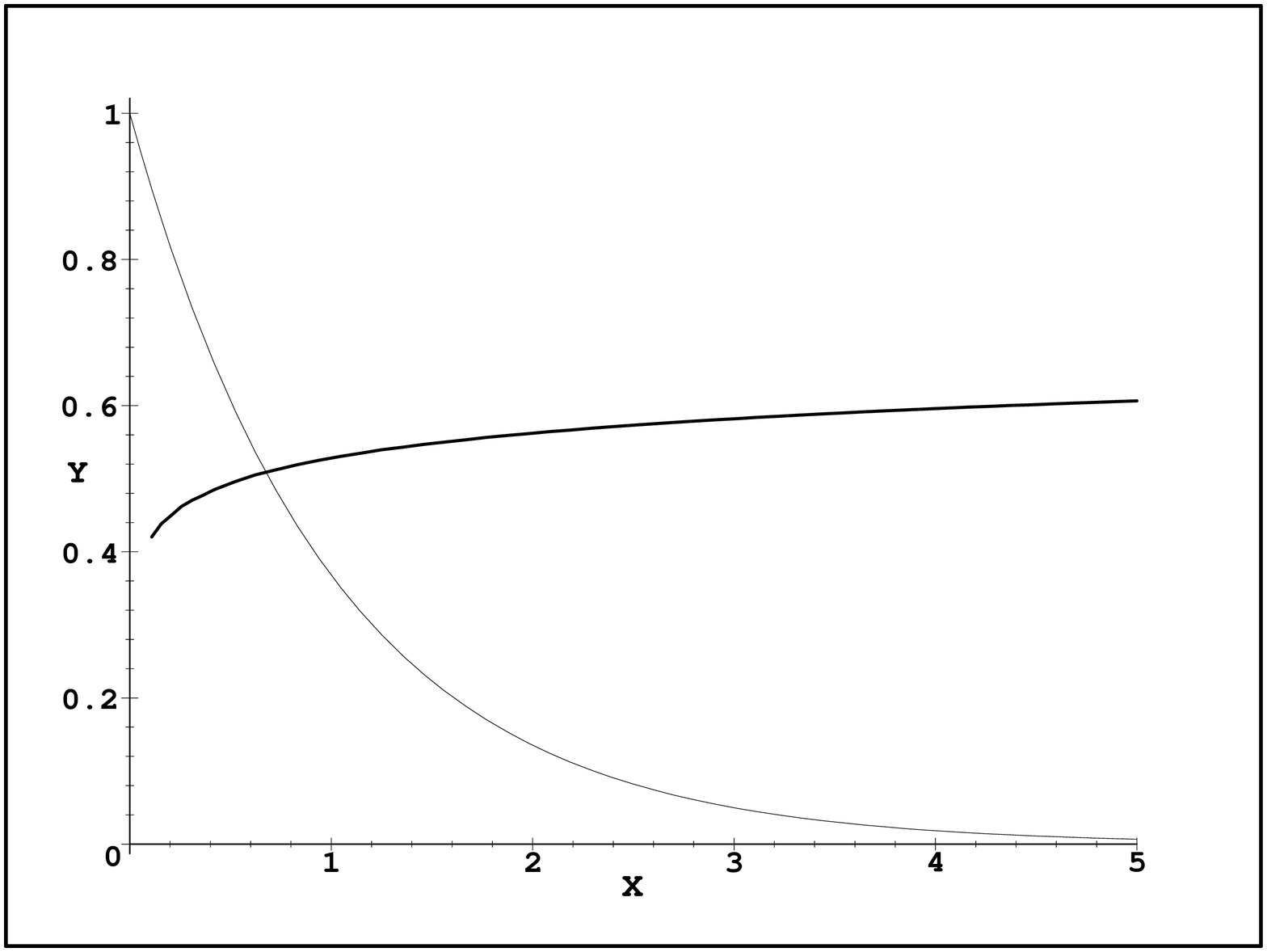}} 
\end{minipage}
\caption{Plot of the function $y=C_{2,\alpha}(x)$ with  
$\alpha=2.2$ is drawn with black color and 
the factor $y=C_{2,0}(x) =\exp (-x)$ is drawn by gray color,
where $x=|{\bf r}|$ and $C(\alpha)/ C^2_e=1$.}
\label{Plot4}
\end{figure}

\begin{figure}[H]
\begin{minipage}[h]{0.47\linewidth}
\resizebox{11cm}{!}{\includegraphics[angle=-90]{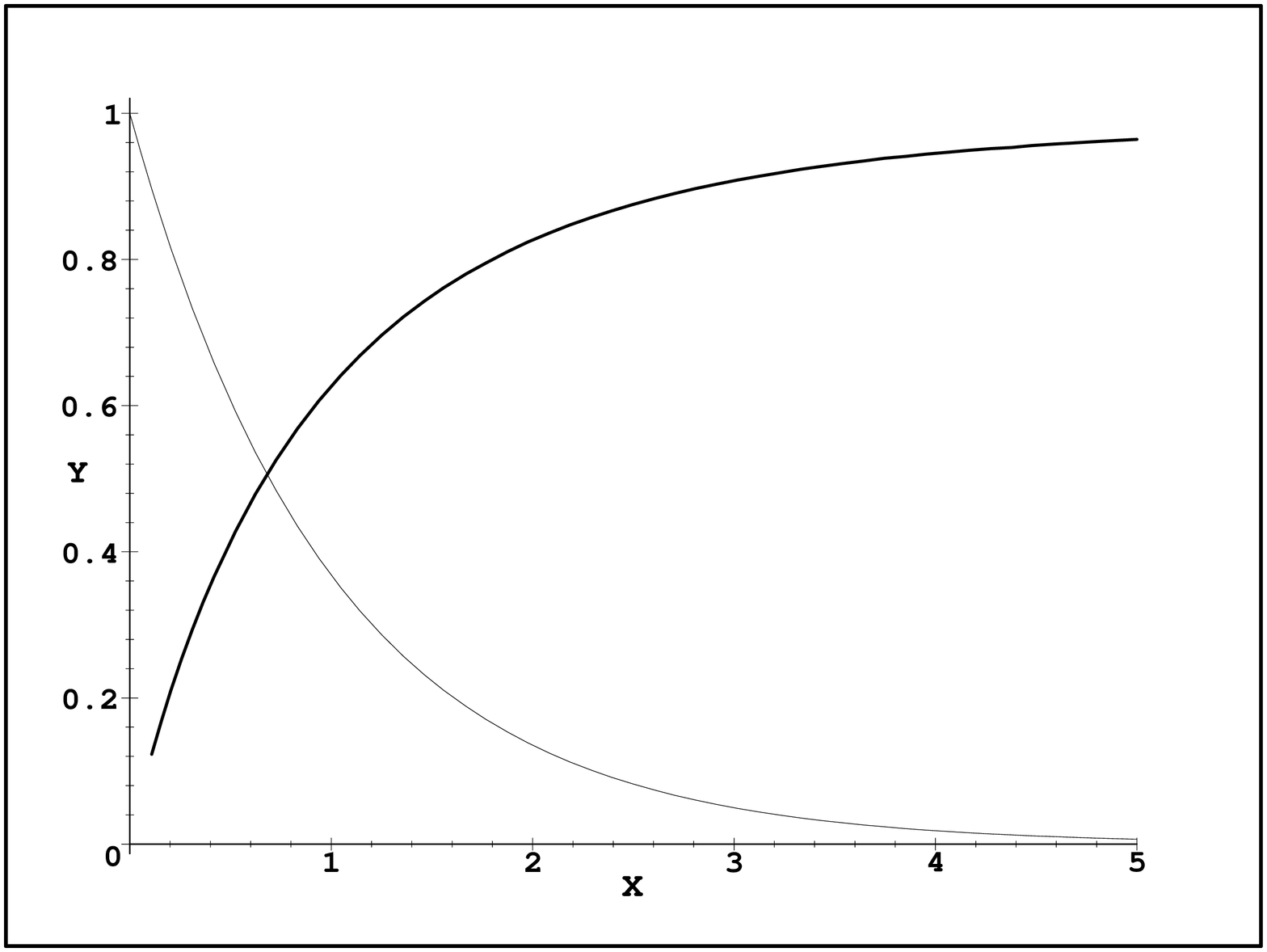}} 
\end{minipage}
\caption{Plot of the function $y=C_{2,\alpha}(x)$ with  
$\alpha=3.6$ is drawn with black color and 
the factor $y=C_{2,0}(x) =\exp (-x)$ is drawn by gray color,
where $x=|{\bf r}|$ and $C(\alpha)/ C^2_e=1$.}
\label{Plot5}
\end{figure}

\begin{figure}[H]
\begin{minipage}[h]{0.47\linewidth}
\resizebox{11cm}{!}{\includegraphics[angle=-90]{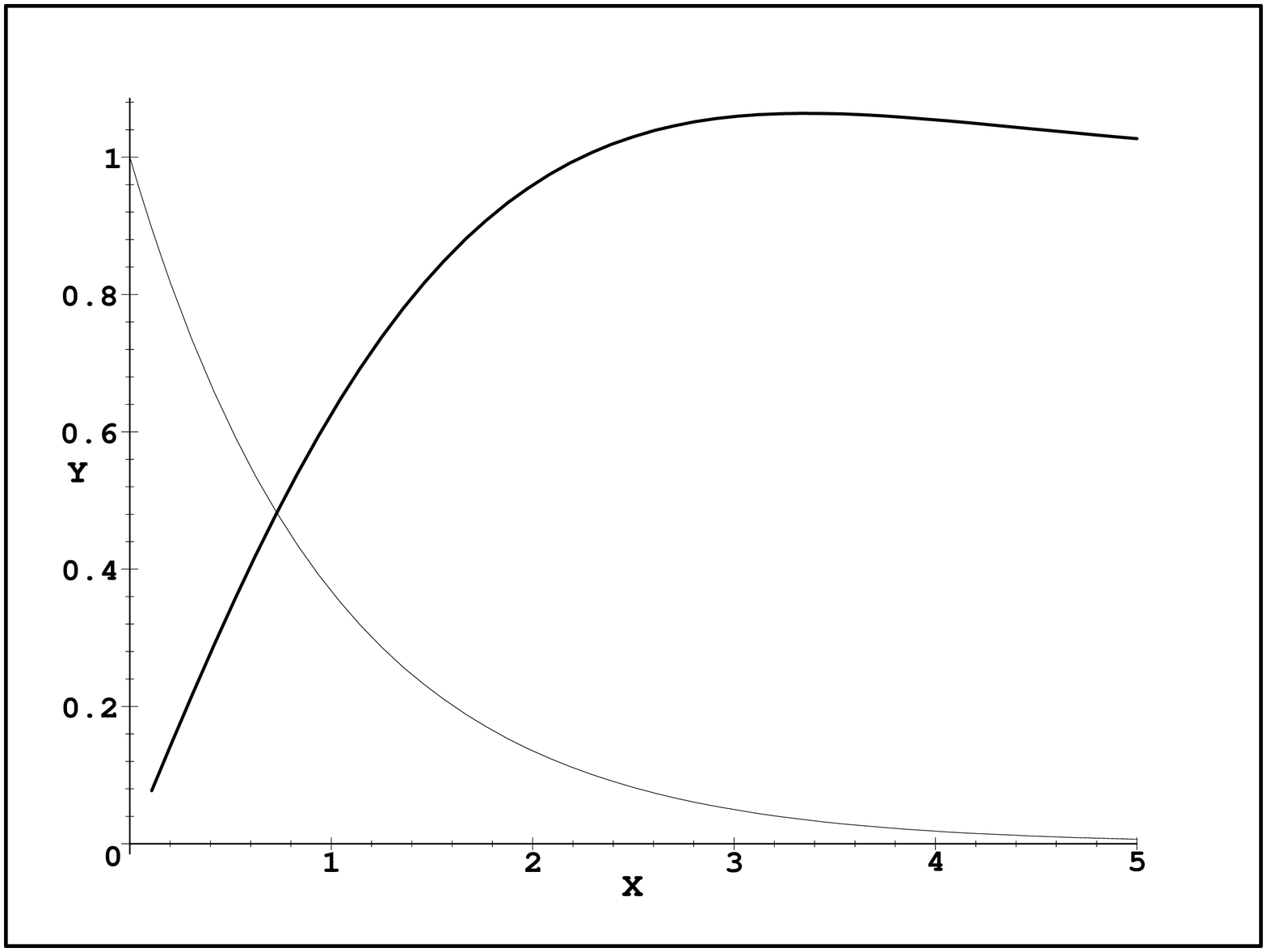}} 
\end{minipage}
\caption{Plot of the function $y=C_{2,\alpha}(x)$ with  
$\alpha=5.9$ is drawn with black color and 
the factor $y=C_{2,0}(x) =\exp (-x)$ is drawn by gray color,
where $x=|{\bf r}|$ and $C(\alpha)/ C^2_e=1$.}
\label{Plot6}
\end{figure}

\section{Conclusion}

A lattice model with long-range interaction of
Gr\"unwald-Letnikov-Riesz type is suggested.
In the continuum limit we derive continuum equations 
with spatial derivatives of non-integer order $\alpha$. 
The correspondent constitutive relations
describe fractional generalization of 
gradient elasticity for $\alpha>2$ and fractional integral elasticity for $0<\alpha <2$. 
The suggested lattice model is considered as a microscopic model 
of the fractional non-local elastic continuum.
We can note that a fractional nonlocal continuum model 
can be obtained from different microscopic or lattice models. 
The benefits of suggested formulation of fractional elasticity are following. 
Firstly, the Gr\"unwald-Letnikov-Riesz derivatives 
in the fractional continuum equations are 
defined by fractional differences.
It can be directly used in numerical simulations of 
fractional gradient and fractional integral elasticity models. 
Secondly, the suggested type of long-range interaction 
for lattice particles allows us to have united lattice model 
for the fractional gradient and fractional integral elasticity. 
We assume that the suggested approach can be generalized 
for 3-dimensional case, for finite strains and plasticity.
An extension of the suggested model for these cases
can be realized by the methods suggested in \cite{TarasovSpringer} (see Sections 8.2, 8.14, 8.15)
with some modifications.
In addition, we note that the model of 
fractional gradient and integral elastic continuum 
has an analog in the plasma-like dielectric material with power-law spatial dispersion \cite{AP2013,POP2013}.
Fractional models of complex material with power-law non-locality allows us to predict unusual properties of materials 
that are characterized by long-range inter-particle interactions. 
These materials can demonstrate a common or universal behavior 
in space by analogy with the universal behavior of low-loss dielectrics in time \cite{Jo1,Jo2,JPCM2008-1,TMF2009}.




\end{document}